\documentclass[11pt]{article}
\usepackage{a4}
\usepackage[dvips]{epsfig}
\usepackage{latexsym}
\usepackage{amssymb}
\usepackage{graphicx}
\usepackage{psfrag}
\def\setb@se#1{\baselineskip=#1 \normalbaselineskip=#1}
\setb@se{14pt}
\def\setb@se#1{\baselineskip=#1 \normalbaselineskip=#1}
\setb@se{14pt}
\newcommand{\be}{\begin{equation}}
\newcommand{\ee}{\end{equation}}
\newcommand{\beqn}{\begin{eqnarray}}
\newcommand{\eeqn}{\end{eqnarray}}
\newcommand{\bsub}{\begin{subeqnarray}}
\newcommand{\esub}{\end{subeqnarray}}

\newcommand{\scr}{\scriptscriptstyle}

\newcounter{subequation}[equation]

\makeatletter \expandafter\let\expandafter\reset@font\csname
reset@font\endcsname
\newenvironment{subeqnarray}
  {\arraycolsep1pt
    \def\@eqnnum\stepcounter##1{\stepcounter{subequation}{\reset@font\rm
      (\theequation\alph{subequation})}}\eqnarray}%
  {\endeqnarray\stepcounter{equation}}
\makeatother

\begin{document}

\title{Spatially Compact Solutions and Stabilization in
Einstein-Yang-Mills-Higgs Theories}
\author{P\'eter Forg\'acs and S\'{e}bastien Reuillon\\
Laboratoire de Math\'{e}matiques et Physique Th\'{e}orique\\
CNRS-UMR 6083\\Universit\'{e} de Tours, Parc de Grandmont\\
37200 Tours, France}

\maketitle

\begin{abstract}
New solutions to the static, spherically symmetric Einstein-Yang-Mills-Higgs equations
with the Higgs field in the triplet resp.\ doublet representation are presented.
They form continuous families parametrized by $\alpha=M_{\rm W}/M_{\rm Pl}$
($M_{\rm W}$ resp.\ $M_{\rm Pl}$ denoting the W-boson resp.\ the Planck mass).
The corresponding spacetimes are regular and have spatially compact sections.
A particularly interesting class with the Yang-Mills amplitude being nodeless
is exhibited and is shown to be linearly {\sl stable} with respect to spherically symmetric
perturbations.
For some solutions with nodes of the Yang-Mills amplitude a new stabilization phenomenon is
found, according to which their unstable modes disappear as $\alpha$ increases (for the triplet)
or decreases (for the doublet).
\end{abstract}

By now a lot is known about classical non-abelian particle-like and black hole solutions
in Einstein-Yang-Mills (EYM) and in Einstein-Yang-Mills-Higgs (EYMH)
theories, we refer to \cite{VolkovGaltsovRep} for a review and references.
The existence of such objects had not been anticipated and their study has revealed
some important non-perturbative features of the interaction between the gravitational and gauge fields.
In addition to particle-like solutions where the corresponding spacetime
is asymptotically flat, a rather different class
has been found in the standard $SU(2)$ EYMH theory (i.e.\ {\sl without a cosmological constant})
with the Higgs field being in the triplet representation \cite{BFM00}.
They form continuous families parametrized by the
gravitational coupling strength, $\alpha=M_{\rm W}/M_{\rm Pl}$,
%(where $M_{\rm W}$ denotes the mass of the gauge field and $M_{\rm Pl}$ stands for the Planck mass)
and the different families are characterized by two integers,
the number of nodes of the Yang-Mills amplitude, $m\geq 1$, resp.\ of the Higgs field, $n\geq 0$.
The spacetimes corresponding to these solutions have
spatially compact sections, i.e.\ they are topologically $\mathbb{R}\times S^3$, representing
 static non-abelian Einstein-universes. 
The solutions of Ref.\ \cite{BFM00} generalize
those found in EYM theory with a {\sl positive cosmological constant} $\Lambda$ (EYM-$\Lambda$)
\cite{Volkovetal96}.
Intuitively the self-interaction potential of the Higgs field acts as a ``dynamical cosmological
constant'' providing the necessary energy density to support a spatially compact spacetime.
The very existence of most of these classical solutions is quite surprising
and their study is important to understand fundamental phenomena in gauge theories
coupled to gravitation.

The important problem of stability of the EYM and EYMH asymptotically flat solutions has been
studied mostly in linear perturbation theory \cite{Volkovetal95,instab},
with the disappointing conclusion that, with the exception of the gravitating analogues of the
't Hooft-Polyakov magnetic monopoles, all of them are unstable.
The stability analysis of the compact solutions in the EYM-$\Lambda$ theory also shows
the existence of unstable modes \cite{Volkovetal296, ForgReuill03}.
Interestingly it has been recently found that,
allowing for a {\sl negative} cosmological constant, the EYM-$\Lambda$ equations
admit nodeless solutions with asymptotically anti-de Sitter (AdS) spacetimes
which have been shown to be {\sl stable} \cite{ADS, ADSstab}.

In this Letter we present globally regular,
static and spatially compact solutions in the
$SU(2)$ EYMH theory with the Higgs field being in the {\sl doublet} representation,
i.e.\ the non-abelian bosonic subsector of the standard model of electroweak interactions.
Similarly as for the triplet Higgs case \cite{BFM00},
there exist distinct continuous families parametrized by $\alpha$ and indexed by the
number of nodes of the YM resp. Higgs field, ($m$,$n$).
We exhibit a new class of compact solutions for both the doublet and triplet case
where the Yang-Mills amplitude {\sl has no nodes}.
These new classes have no analogues in EYM-$\Lambda$ 
theories and their existence is related to an interplay of the scalar potential
and the gravitational field. They could also be interpreted as gravitationally bound
(or rather confined) monopole-antimonopole (triplet) or sphaleron-antisphaleron (doublet) pairs.

We have also investigated the stability properties of the compact solutions.
The spectrum of small fluctuations can be divided with respect to parity
into even ("gravitational") and odd sectors.
A detailed numerical investigation strongly indicates linear {\sl stability} of these
nodeless solutions, at least against spherically symmetric perturbations
for both the triplet and the doublet case in the ``gravitational sector'' of perturbations
which makes their further study worthwhile.
%Amazingly these nodeless solutions show no gravitational instabilities which plague the
%previously known ones.
In the odd parity sector of perturbations an analytic argument
shows that for the triplet case
the number of unstable modes is simply equal to the number of nodes, $m$, of the Yang-Mills
field, establishing that the nodeless solutions have {\sl no unstable modes at all}.
In the doublet case, however, a suitable adaptation of the result of Ref.\
\cite{Volkov-sphaleron}
leads to the result that there is always an unstable mode
independently of $m$. This unstable mode does not couple directly
to the metric fluctuations and it might just
indicate that the {\sl static} doublet solution becomes time-dependent,
possibly of oscillatory type since 3-space is compact and there are no radiative modes
as in an asymptotically flat space-time.

In addition, we have found yet another interesting phenomenon --``stabilization'', which occurs
in the even parity sector.
Our stability analysis has shown that for the triplet case,
the $(m=1,\,n=0)$ family of compact solutions
discovered in \cite{BFM00} has two unstable modes in the gravitational sector.
Quite surprisingly
as $\alpha$ increases the two unstable modes actually {\sl merge}, and for larger values of
$\alpha$ no unstable modes are found.
This phenomenon can be interpreted so that the solutions undergo
``gravitational stabilization'' since they loose their unstable modes for {\sl increasing}
values of the gravitational coupling strength.
To the best of our knowledge this is the first example
that a system gets stabilized when the gravitational coupling gets stronger.
In fact exploiting the (somewhat unphysical) limit $\alpha\to\infty$ we have strong numerical
evidence that no instabilities occur for arbitrarily strong values of the gravitational coupling.
Remarkably there is also a family of solutions $(m=2,\,n=0)$ in the doublet case
which looses its two unstable modes in the gravitational sector of perturbations.
This occurs, however, for {\sl decreasing} values of $\alpha$.
The classical solutions we find exist in the regime of strong gravitational
coupling -- at energies close the Planck scale, and therefore they would be mostly relevant
for semiclassical computations and to gain more understanding of the classical phase space
of EYMH theories. They seem to be quite distinguished by their stability properties and also
by the fact that they are present in the standard EYMH theories
(i.e.\ without the introduction of new interactions, or negative vacuum energies)
making them of quite some interest.

The most general spherically symmetric line element is
\be\label{ansatzmetric}
ds^2=e^{2\nu(R,t)}dt^2-e^{2\lambda(R,t)}dR^2-r^2(R,t)d\Omega^2 \,,
\ee
and the most general spherically symmetric Ans\"atze for the gauge
resp.\ for the Higgs fields can be written as
\bsub\label{YMHansatze}
A &=& (A_{\scr 0} dt+A_{\scr 1} dR + \cos\theta d\phi)\,T_{\scr 3}+
(W\,T_{\scr 1}+K\,T_{\scr 2})\,d\theta+(W\,T_{\scr 2}-K\,T_{\scr 1})\,\sin\theta d\phi\,,\\
\Phi &=& H\,.\,\mathbf{1} + i\;F\;T_{\scr 3}\ \hbox{(doublet)}\,, \quad
\Phi = H\; T_{\scr 3}\ \hbox{(triplet)}
\esub
where $T_a\,(a=1,2,3)$ are the generators of $SU(2)$ and $A_i,W,K,H,F$
are functions of $R$ and $t$.
To look for compact solutions we use the ``minimal'' Ansatz  compatible with the field equations,
corresponding to $A_{\scr 0}=A_{\scr 1}=K=F=0$.
Following the notations of \cite{BFM00}, we define
$N:=\dot r$, $\kappa:=N+r\dot{\nu}$,
where $\dot{f}:={d f/ d\sigma}:= e^{-\lambda}df/dR$,
and the static field equations can be written as
\bsub\label{stateqs}
\dot{N}&=&\frac{\kappa-N}{r}N-2\frac{\dot{W}^2}{r}-d\,r\dot{H}^2\,,\\
\dot{\kappa}&=&[1-\kappa^2+2\dot{W}^2-\frac{\beta^2r^2}{2}(H^2-\alpha^2)^2-2C]/r\,,\\
\ddot{W}&=&W\frac{W^2-1}{r^2}+\frac{1}{2}\frac{\partial C}{\partial W}
-\frac{\kappa-N}{r}\dot{W}\,,\\
\ddot{H}&=&\frac{1}{d}\left(\frac{\beta^2}{2}H(H^2-\alpha^2)
+\frac{1}{r^2}\frac{\partial C}{\partial H}\right)
-\frac{\kappa+N}{r}\dot{H}\,,\
\esub
where $\beta=M_{H}/M_{W}$, $M_{H}$ being the mass of the Higgs field,
and $d=2$, $C=(W-1)^2H^2/2$ resp.\ $d=1$, $C=W^2H^2$
for the doublet resp.\  for the triplet case. It should be noted that for the doublet
case $\alpha$ is $M_{\rm W}/\sqrt{2}M_{\rm Pl}$.
Additionally there is a constraint equation relating
\{$r$,$N$,$\kappa$,$H$,$W$\} to the first derivatives of the matter fields.

Eqs.~(\ref{stateqs}) admit a family of local solutions regular at the origin
where $r=\sigma+O(\sigma^3)$ and $N=\kappa=1+O(\sigma^2)$.
For the triplet case, the behaviour of the YM resp.\ Higgs field at the origin is 
$W=\pm(1-b\,\sigma^2)+O(\sigma^4)$ resp.\ $H=a\,\sigma+O(\sigma^3)$
with $a$, $b$ being free parameters. One can impose without restriction $W(0)=+1$.
For the doublet case, however,
because of the asymmetry of the coupling term,
$C$, between the Higgs and YM fields with respect to $W\to-W$,
there are two distinct
families of local solutions corresponding to the sign of $W(0)$:
either $W=-(1-b\,\sigma^2)+O(\sigma^4)$ and $H=a\,\sigma+O(\sigma^3)$,
or $W=+(1-b\,\sigma^2)+O(\sigma^4)$ and $H=a+O(\sigma^2)$.

For $H(\sigma)\equiv 0$ the EYMH eqs.\
reduce to those of an EYM-$\Lambda$ theory with the cosmological constant
given as
$\Lambda=\alpha^4\beta^2/4$. As found in Ref.\ \cite{Volkovetal96}
for a special set of values of the cosmological constant $\{\Lambda=\Lambda_m, m=1\ldots\}$
the EYM-$\Lambda$ theory admits regular compact solutions.
 The $m=1$ solution is known analytically \cite{AnalSol}:
\be\label{analytsol}
r=\sqrt{2}\sin(\sigma/\sqrt{2})\,,\;\;
N=\kappa=W=\cos(\sigma/\sqrt{2})\,,
\ee
with $\Lambda_1=3/4$. Due to the compactness of the manifold
($S^3$ in this case) there is an extremum of the radial function,
$\dot r(\sigma_e)=0$, to which we refer as an equator.
Following Ref.\ \cite{BFM00} a natural strategy to search for solutions of
Eqs.\ (\ref{stateqs}) is to look for solutions bifurcating with those of
the EYM-$\Lambda$ theory.
At the bifurcation point the bifurcating solutions are normalizable zero modes
of the linearized Higgs field
Eq.\ (\ref{stateqs}d) in the corresponding EYM-$\Lambda_m$ background solution.
In the doublet case the equation for the linearized Higgs field amplitude,
$h$, takes the form:
\be\label{bifucation}
\ddot{h}=\left(\frac{(W_{\rm b}-1)^2}{2r_{\rm b}^2}-\frac{\Lambda_{m}}{\alpha^2}\right)h
-\frac{\kappa_{\rm b}+N_{\rm b}}{r_{\rm b}}\dot{h}\,,
\ee
where $r_{\rm b}\,,N_{\rm b}\ldots$ etc.\ stand for the background solution.
The bifurcation parameter, $\alpha$, is to be varied to find globally regular
solutions of Eq.\ (\ref{bifucation}).
In the analytically known $m=1$ background (\ref{analytsol})
the bounded zero modes of the Higgs-field, $h_n$, can be explicitly constructed.
They are characterized by an additional node number, $n\geq 0$,
and by the values of the bifurcation parameter
$\alpha_{\scr 1,n}^2={6}/({(2n+1)(2n+5)-2})\,,\quad n=0,1,\ldots$,
(yielding $\alpha^2_{\scr 1,0}=2$, $\alpha^2_{\scr 1,1}=6/19$, \dots).
Due to the asymmetric form of $C$ for the doublet, $h_n$ is neither symmetric
nor antisymmetric around the equator, and the solutions of the full system
are {\sl asymmetric} for odd values of the Yang-Mills node number.
For odd values of $m$ the gauge field amplitude interpolates between
$W=+1$ (at the north pole) and $W=-1$ (at the south pole) of the
compact manifold.
For even values of $m$, there are two classes of solutions,
both symmetric around the equator distinguished by the values of
$W(0)=\pm 1$ at the origin (referred to as type $+$ resp.\ $-$).
We give a detailed description of these families in \cite{ForgReuillinprep},
here we exhibit a particularly interesting class of type $W(0)=-1$
with $m=2$ and $n=0$, denoted as $(m=2,\,n=0)^-$.
This solution bifurcates with the $m=2$ compact solution of the
EYM-$\Lambda$ theory at $\alpha_{\scr 2,0}=1.494277$ with
$\Lambda_2=0.364244$.
It exists for a finite range of values of 
$\alpha\in[\alpha_{\scr \rm min},\alpha_{\scr 2,0}]$ with $\alpha_{\scr \rm min}\approx0.54(9)$. 
For decreasing values of $\alpha$, the two zeros of $W$ move
towards the equator and they merge there for $\alpha_{\star}=0.807469$.
For $\alpha<\alpha_{\star}$, these solutions become nodeless.
One can easily show that this cannot happen for the $(m=2,n=0)^+$ solutions.
For $\alpha\to\alpha_{\scr \rm min}$ the spatial sections of the limiting solution
will no longer be compact ($\sigma\to\infty$).

\begin{figure}[ht]
\hbox to\linewidth{\hss%
  \psfrag{X}{$\alpha=0.89082\;\beta=8.29424$}
  \psfrag{sigma}{$\sigma$}
  \psfrag{W,H}{{\scriptsize $W,H$}}
  \psfrag{N}{{\scriptsize $N$}}
  \psfrag{r,k}{{\scriptsize $r,\kappa$}}
    \resizebox{10cm}{7cm}{\includegraphics{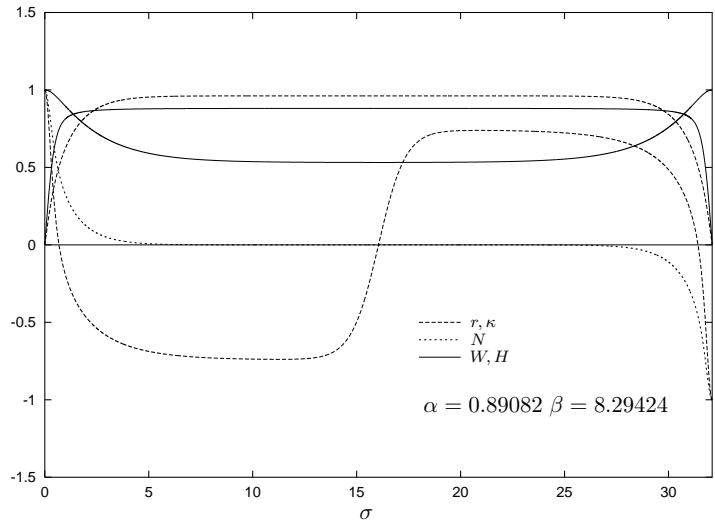}}%
    \hss}
\caption{\label{plotbackt}A nodeless triplet solution for $\alpha$ close to $\alpha_{\scr 0}$.}
\end{figure}

\begin{figure}[ht]
\hbox to\linewidth{\hss%
  \psfrag{X}{$\alpha=0.7\;\beta=5.45095$}
  \psfrag{sigma}{$\sigma$}
  \psfrag{W,H}{{\scriptsize $W,H$}}
  \psfrag{N}{{\scriptsize $N$}}
  \psfrag{r,k}{{\scriptsize $r,\kappa$}}
   \resizebox{10cm}{7cm}{\includegraphics{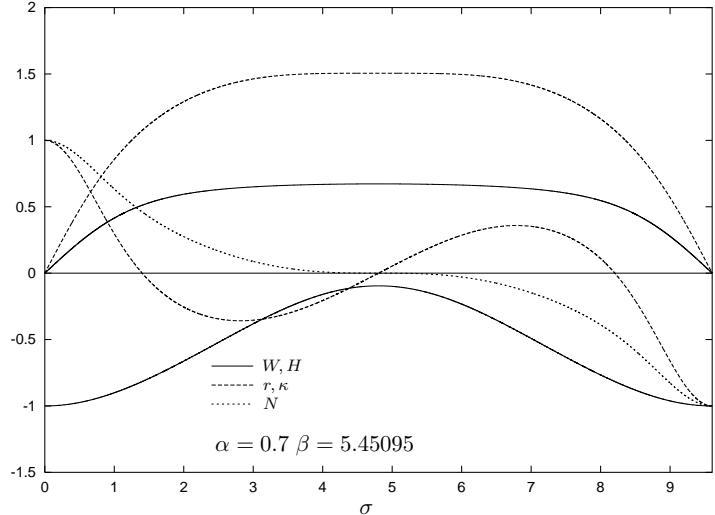}}%
\hss} \caption{\label{plotbackd}A nodeless doublet solution.}
\end{figure}

For the triplet case, the corresponding analysis has been performed in
\cite{BFM00}
where various one parameter families of solutions of
Eqs.~(\ref{stateqs}), indexed by the node numbers $m\geq1$, $n\geq0$,
have been found.
These solutions are symmetric or antisymmetric around
the equator depending on the parity of $m$ and $n$.
Here we exhibit a new type of family of solutions in the triplet case
with $m=0$, $n=0$.
This family does not bifurcate with any known EYM-$\Lambda$
compact solution and it seems to exist for $\alpha\in[\alpha_{\scr 0},\infty)$
where $\alpha_{\scr 0}\approx0.87(7)$. 
In the strong coupling limit $\alpha\to\infty$,
Eqs.\ (\ref{stateqs}) reduce to those of an EYMH system with
$\beta=0$ and with a cosmological constant $\Lambda=1/4$.
In this limit the equator, and hence the zero of $W$, tend to infinity,
and the spatial sections of the solutions will no longer be compact
($\sigma\to\infty$ and $r\to\sqrt{2}$).
The limiting metric has the asymptotic form
$ds^2=\sigma^2\,dt^2-d\sigma^2-2\,d\Omega^2$,
corresponding to the product of a 2-d Minkowski space
(in Rindler coordinates) and a 2-sphere.
In the limit $\alpha\to\alpha_{\scr 0}$
the solution interpolates between a regular origin and a non-abelian
fixed point of Eqs.\ (\ref{stateqs}) (see Ref. \cite{BFM2}) with non-compact
spatial sections.
The nodeless solutions for both the triplet and doublet
cases are displayed on Figs.\ \ref{plotbackt}, \ref{plotbackd}.

{\bfseries Stability Analysis}.
We shall restrict our stability analysis to spherically symmetric
fluctuations around the solutions. In many cases instabilities show up
already in the spherically symmetric sector. The complete
stability analysis of the asymptotically AdS EYM-$\Lambda$ solutions
performed in Ref.\ \cite{ADSstab} has demonstrated that solutions stable
under spherically symmetric perturbations remain stable under non-spherically
symmetric perturbations.
The general stability analysis is considerably more involved,
and it is further complicated for spatially compact solutions
due the existence of an equator.
The most general spherically symmetric fluctuations
split into two sectors according to spatial parity reflections.
Making a mode decomposition of the type
$f(R,t)=f(R)+\delta f(R)\,e^{i\omega t}$,
where $f(R)$ denotes a static solution, the linearized EYMH Eqs.\
fall into two sets. In the even
parity ("gravitational") sector,
the matter perturbations ($\delta W$, $\delta H$)
couple to the gravitational ones ($\delta\nu$, $\delta\lambda$, $\delta r$).
The odd parity ("sphaleronic") sector involves
only the matter perturbations $\delta A_i$, $\delta K$ and $\delta F$
(for the doublet).

In the even parity sector, there exists a particular gauge --
the Schwarzschild gauge -- defined by $\delta r=0$,
in which the metric perturbations can be eliminated
and one obtains
a closed system of Schr\"odinger equations for the
perturbation amplitudes $\Phi:=(\delta W\,,\delta H)$ of the form:
\be\label{2channel}
e^\nu(e^\nu\dot\Phi)^{\displaystyle\bf \cdot}=(V+\omega^2)\Phi\,,
\ee
where the $2\times2$ potential matrix, $V$, can be expressed explicitly
in terms of the functions $\{r,N,e^\nu,H,W\}$ and their derivatives.
For the limiting solutions ($\alpha\to\infty$)
%with the asymptotic (\ref{bcinf}),
the potential matrix is bounded and one can use a generalized version of the
Jacobi criterion \cite{AmannQuittner95} to detect bound states with negative
eigenvalues ($\omega^2<0$), corresponding to unstable modes.
According to this criterion, the vanishing of a suitable determinant
of the solutions of Eqs.\ (\ref{2channel}) in $(0,+\infty)$ implies the
existence of at least one bound state with a negative eigenvalue.
We have numerically integrated (\ref{2channel}) in the background of the limiting
$(m=0,\,n=0)$ solution and found that it has no unstable mode.
\begin{figure}[ht]
\hbox to\linewidth{\hss%
  \psfrag{XX}{}
  \psfrag{x}{$\alpha_{\scr 1,0}$}
  \psfrag{alpha}{$\alpha$}
  \psfrag{om0}{$\omega_{\scr 0}^2$}
  \psfrag{om1}{$\omega_{\scr 1}^2$}
    \resizebox{10cm}{7cm}{\includegraphics{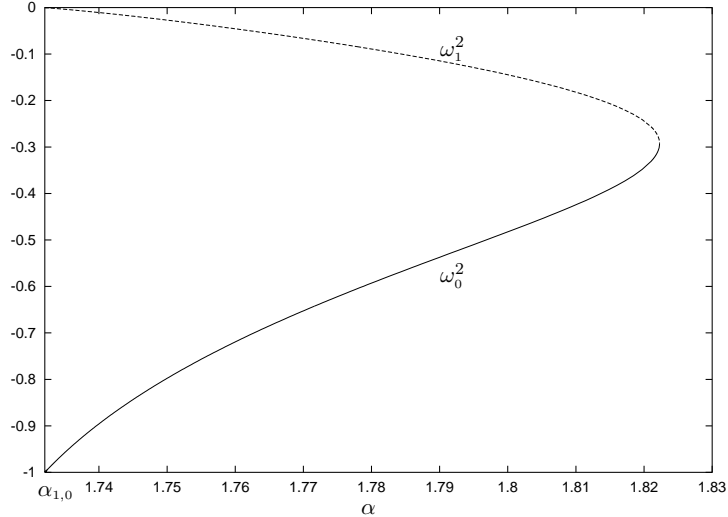}}%
 \hss}
\caption{\label{plotfreq1}Stabilization for the $(m=1,\,n=0)$ triplet solution. The bifurcation point is $\alpha_{\scr 1,0}=1.732051$}
\end{figure}
\begin{figure}[ht]
\hbox to\linewidth{\hss%
  \psfrag{XX}{}
  \psfrag{x}{$\alpha_{\scr 2,0}$}
  \psfrag{alpha}{$\alpha$}
  \psfrag{om0}{${\omega_{\scr 0}^2}$}
  \psfrag{om1}{${\omega_{\scr 1}^2}$}
  \psfrag{om2}{${\omega_{\scr 2}^2}$}
    \resizebox{10cm}{7cm}{\includegraphics{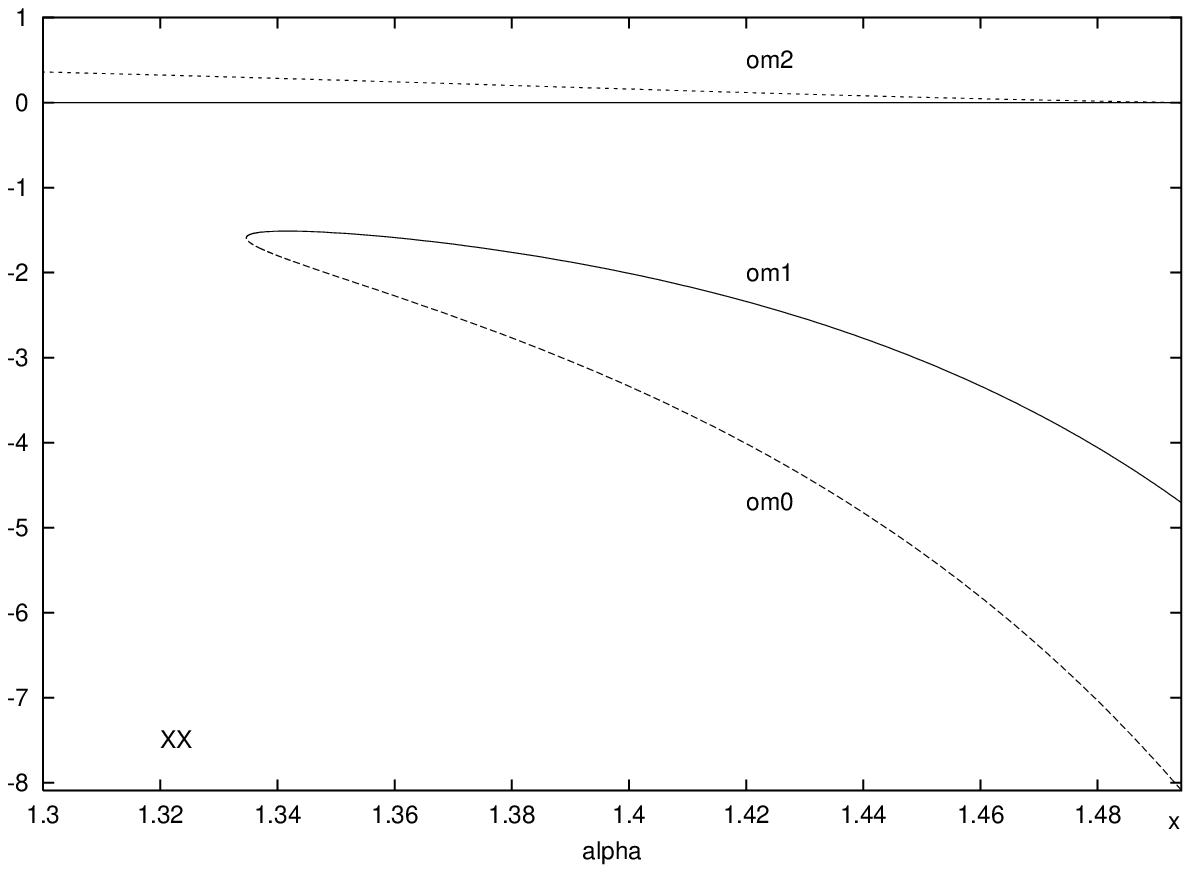}}%
\hss} \caption{\label{plotfreq2}Stabilization for the
$(m=2,\,n=0)^-$ doublet solution. The bifurcation point is $\alpha_{\scr 2,0}=1.494277$}
\end{figure}

For solutions with compact spatial sections (i.e.\ topologically $S^3$),
the potential matrix, $V$, in Eqs.~(\ref{2channel})
is {\sl singular due to the presence of an equator}, see Refs.\ \cite{Volkovetal296,ForgReuill03}.
%Indeed, the $\delta r=0$ gauge is so singular that some physically admissible eigenfunctions
%become not even square integrable,
One can avoid such problems by working in
a {\sl globally regular} gauge, e.g.\ $\delta\lambda=0$.
In this gauge, however, one obtains a somewhat unusual eigenvalue problem,
which can be nevertheless dealt with numerically.
Starting at the bifurcation points where the
numerical values of the negative eigenvalues are known
\cite{ForgReuill03} it is possible to determine their dependence on $\alpha$.
We remark that the zero mode of the Higgs field
at the bifurcation points, becomes an unstable mode for the
($m=1$, $n=0$) triplet solution, comp.\ Fig.\ \ref{plotfreq1}, while
it goes into the positive spectrum for the
$(m=2,\;n=0)^-$ doublet solution, comp.\ Fig.\ \ref{plotfreq2}.
The negative eigenvalues behave quite
similarly to what has already been observed for certain
families in the EYM-$\Lambda$ system \cite{ForgReuill03}.
It is quite remarkable that in the present cases
the solutions loose {\sl all} of their unstable modes.

For the nodeless triplet solutions, we have undertaken a careful
numerical search for unstable modes (scanning in the parameter space)
and the lack of any evidence for such modes, together with the result
of the Jacobi criterion in the $\alpha\to\infty$ limit
strongly suggests that this solution has no unstable modes for any
$\alpha\in[\alpha_0,\infty)$.

In the odd parity sector, the stability analysis is simplified by the fact that the
metric and matter perturbations decouple.
By a suitable adaptation of the main ideas of
Refs.~\cite{Volkovetal95,Volkovetal296}
we have proved the following:
a) in the triplet case, solutions with $m$ nodes for the
YM amplitude have exactly $m$ unstable modes;
b)in the doublet case, all solutions have at least one
unstable mode.

Therefore we conclude that the nodeless triplet solutions are
linearly {\sl stable} against
spherically symmetric perturbations in both sectors,
while the $(m=1,\,n=0)$ triplet solutions stabilize in the even parity
sector but still possess one unstable mode in the odd
parity sector.\\
The $(m=2,\;n=0)^-$ doublet solutions are stable for $\alpha\lesssim 1.334$
in the even parity sector
and have at least one unstable mode in the odd parity sector. 
It is not unlikely that the gravitational stability persists at the non linear level,
and no singularity forms.
Then, the instability of the static solution indicates
merely that it becomes time-dependent (of oscillatory type).
\end{document}